\documentclass[aip,reprint]{revtex4-1}

\usepackage{graphicx}
\usepackage{epsfig}
\usepackage{amsmath,amssymb}
\usepackage{bbm}
\usepackage{dcolumn}
\usepackage{bm}
\usepackage{textgreek}
\usepackage[title]{appendix}
\numberwithin{equation}{section}

\begin{document}

\def\bfr{{\bf r}}
\def\bfR{{\bf R}}

\preprint{APS/123-QED}

\title{Closure to the PRISM equation derived from nonlinear response theory}
\author{James P. Donley}
\email{jpdonley7@icloud.com}
\affiliation{Material Science Institute, University of Oregon, Eugene, OR 97403, USA}
\date{\today}
\begin{abstract}
Nonlinear response theory is employed to derive a closure to the polymer reference interaction site model (PRISM) equation. The closure applies to a liquid of neutral polymers at melt densities. It can be considered a molecular generalization of the mean spherical approximation (MSA) closure of Lebowitz and Percus to the atomic Ornstein-Zernike (OZ) equation, and is similar in some aspects to the reference ``molecular'' MSA (R-MMSA) closure of Schweizer and Yethiraj to PRISM.  For a model binary blend of freely-jointed chains, the new closure predicts an unmixing critical temperature, $T_c$, via the susceptibility route that scales linearly with molecular weight, $N$, in agreement with Flory theory. Predictions for $T_c$ of the new closure differ greatest from those of the R-MMSA at intermediate $N$, the latter being about 40\% higher than the former there, but at large $N$ both theories give about the same values. For an isotopic blend of polyethylene, the new and R-MMSA closures predict a $T_c$ about 25\% higher than the experimental value, which is only moderately less accurate than the prediction of atomic OZ-MSA theory for $T_c$ of methane. In this way, the derivation and its consequences help to identify the ingredients in a theory needed to model properly the equilibrium properties of a polymeric liquid at both short and long lengthscales.
\end{abstract}


\maketitle

\section{Introduction}
There are at least five generalizations of the Ornstein-Zernike (OZ) theory of homogeneous, equilibrium atomic liquids\cite{Hansen1986} to molecules: molecular Ornstein-Zernike theory;\cite{Blum1972a,Ishizuka2012} the reference interaction site model (RISM)\cite{Chandler1972,Chandler1982} and its specialization to polymers (polymer-RISM or PRISM);\cite{Curro1987,Schweizer1997} the diagrammatically proper integral equation theory\cite{CSL82} and the (at least in one case) formally equivalent optimized cluster theory;\cite{Lupkowski1987,Melenkevitz1997} the Wertheim associating fluid theory;\cite{Wertheim1984,KR93} and two-molecule theory.\cite{Laria1991,DCM94} By far the most popular though, perhaps from its ease of use,\cite{Pyprism2018} has been RISM/PRISM. 

However, as has been shown,\cite{Chandler1976} RISM is not diagrammatically proper. As a consequence there has been up to now only a few formally derived closures to it. In one approach, Chandler employed a functional derivative scheme to derive a closure to RISM.\cite{Chandler1973} While this closure has been shown to be accurate primarily at low density,\cite{Takebayashi1996} its form was used\cite{Chandler1973} to justify interpreting RISM as an optimized random phase approximation (RPA) theory, i.e., identifying the atomic mean spherical approximation (MSA)\cite{Lebowitz66} as an accurate closure. A later analysis of a model liquid of thread polymers using a Gaussian field theory concluded similarly.\cite{Chandler1993} In another approach done in two-molecule theory, classical density functional theory\cite{CMS86a} was used to derive an equation for the radial distribution function. While strictly a separate theory, the medium-induced potential appearing in the two-molecule equation can be recast to connect to RISM.\cite{DCM94,Tanaka2014} Efficient algorithms to solve this theory have not been developed until recently though.\cite{Li2022} In practical use of RISM then, choosing a particular closure has tended to be more empirical than that done for atomic liquids with OZ theory.

In a previous work,\cite{Donley05b} referred hereafter as (I), a self-consistent equation for the pair radial distribution function was derived using a molecular density functional theory for classical liquids. The general properties of  polymer blends and diblock copolymer melts were analyzed. The theory was shown to predict properly the molecular weight dependence of the blend critical temperature, and the disordered to ordered lamellar transition in diblocks. However, the manner of enforcing the core condition on the radial distribution function was not as strong as that commonly done in the OZ class of theories.

In this work, nonlinear response theory is used to derive more simply the basic equations obtained in (I). Then making approximations appropriate to liquids at melt densities with nonbonded potentials that consist of a strongly repulsive, short-range core, and a weak, long-range tail, such as a Lennard-Jones, a closure to the PRISM equation results. In that way, the closure is derived in an indirect manner, similar to that done in the derivations in two-molecule theory, but more deliberately. Evidence is presented that this closure for polymeric liquids is almost as accurate as the MSA closure for atomic liquids.

\section{\label{sec:theory}Theory}
\subsection{System definition}
It is convenient to work in the grand canonical ensemble. Consider then a liquid of identical molecules in equilibrium in a fixed volume $V$, in strong contact with an external bath at temperature $T$, and in strong contact with an external source of molecules at chemical potential $\mu$. This potential sets the average number of system molecules at $M$, giving an average molecular density of $\rho=M/V$. Each molecule is composed of $n_t$ types of spherical sites, with $N_k$ sites of type $k$. The average density of type-$k$ sites is then $\rho_k=N_k\rho$. Let the sites on a molecule be labelled by $\alpha = 1, 2,\cdots, N$, where $N$ is the total number of sites per molecule. Further, let the position of the $\alpha$th site on the $i$th molecule be denoted by $\bfr_{\emph{i} \alpha}$. Unless explicitly stated otherwise, in what follows all energies will be expressed in units of $\emph{k}_BT$, where $\emph{k}_B$ is Boltzmann's constant. 

Of particular interest is the density-density correlation function between a site of type $k$ at position ${\bfr}$ and one of type $k'$ at position $\bfr'$:
\begin{equation}
  S_{kk'}({\bfr},{\bfr}') = \sum_{\substack{\alpha\in k \\ \beta\in k'}}^{N} S_{\alpha\beta}(\bfr,\bfr').
\label{eq:Sofrkk}
\end{equation}
where $S_{\alpha\beta}(\bfr,\bfr')$ is the density-density correlation function between the $\alpha$th site on a molecule at $\bfr$ and the $\beta$th site on the same or different molecule at $\bfr'$. It is defined as
\begin{equation}
 S_{\alpha\beta}({\bfr},{\bfr}') = \langle {\hat\rho}_{\alpha}({\bfr}) {\hat\rho}_{\beta}({\bfr}')\rangle - 
                             \langle {\hat\rho}_{\alpha}({\bfr})\rangle \langle{\hat\rho}_{\beta}({\bfr}^\prime)\rangle,
\label{eq:Sofr}
\end{equation}
where
\begin{equation}
 {\hat \rho}_{\alpha}(\bfr) = \sum_{\emph{i}=1}^M
                             \delta \left(\bfr-\bfr_{\emph{i} \alpha} \right)
\end{equation}
is the microscopic density of sites $\alpha$ at point $\bfr$, with $\delta(\bfr)$ being the Dirac delta function. The brackets denote thermodynamic averages. If the liquid is homogeneous and isotropic, $S_{kk'}({\bfr},{\bfr}') \rightarrow S_{kk'}(r)$, with $r = \vert {\bfr} - {\bfr}' \vert$.
The Fourier transform of $S_{kk'}(r)$ is the partial structure factor, ${\hat S}_{kk'}(q)$, where $q\equiv\vert\bf q\vert$ is the wavevector conjugate to $r$.

It is convenient to represent $S_{kk'}(r)$ as the sum of intra- and intermolecular contributions
\begin{equation}
S_{kk'}(r) = \Omega_{kk'}(r) + H_{kk'}(r).
\label{Sofrdef}
\end{equation}
The intramolecular correlation function
\begin{equation}
  \Omega_{kk'}(r) = \rho\sum_{\substack{ \alpha \in k \\ \beta\in k'}}^{N}\omega_{\alpha\beta}(r),
\label{omegadef}
\end{equation}
where
\begin{equation}
\omega_{\alpha\beta}(r) = \frac{1}{M}\sum_{i=1}^M \left\langle \delta(\textbf{r}-\textbf{r}_{i\alpha}+\textbf{r}_{i\beta})\right\rangle
\label{omegaalfbet}
\end{equation}
is the probability density of sites $\alpha$ and $\beta$ on the same molecule being a distance $r$ apart.
The intermolecular correlation function 
\begin{equation}
  H_{kk'}(r)= \rho_k \rho_{k'}h_{kk'}(r) = \rho_k \rho_{k'}[g_{kk'}(r)-1],
\label{Hofrdef}
\end{equation}
where 
\begin{equation}
 g_{kk'}(r)=\frac {1}{N_kN_{k'}}\sum_{\substack{ \alpha \in k \\ \beta\in k'}}^{N} g_{\alpha\beta}(r),
\label{eq:gofr}
\end{equation}
is the site averaged form of 
\begin{equation}
 g_{\alpha\beta}(r)= \frac{1}{M(M-1)}\sum_{i\neq j}^M  \left  \langle V\delta (\bfr-\bfr_{i \alpha}+\bfr_{j\beta})\right \rangle,
\label{eq:gofralfbet}
\end{equation}
which is the radial distribution function between sites of type $\alpha$ and $\beta$ on different molecules.

With $g_{kk'}(r)$ and ${\hat S}_{kk'}(q)$ one can compute thermodynamic quantities such as the pressure, the intermolecular contribution to the internal energy, and portions of the phase diagram, which is the goal here.  The partial structure factors, being proportional to the scattered intensity in the single scattering limit, can also be compared directly with experiment.

\subsection{Nonlinear response to an external field}
The essence of the approach here is to use the standard statistical mechanical formalism for classical molecules to derive an expression for the equilibrium density response of the system to an externally imposed field. This density response is then related to the radial distribution function of a homogeneous liquid, and from that Eqs.~\eqref{Sofrdef}-\eqref{Hofrdef} are used to obtain the partial structure factors.

To that end, impose an external field $\phi$ on the liquid. The field depends on the molecule site and can vary spatially, and so has components $\phi_\alpha({\bfr})$. Further, instead of having a fixed value, let the field strength vary with a parameter $\lambda$, so that $\phi_\alpha({\bfr},\lambda=0) = 0$, and $\phi_\alpha({\bfr},\lambda = 1) = \phi_\alpha({\bfr})$, its full value. In the following derivation, quantities will be assumed to vary continuously with $\lambda$. In the theory of classical liquids, this procedure is a variation on the thermodynamic integration, i.e., ``charging'' technique,\cite{Hill1987} and goes back at least to Kirkwood.\cite{Kirkwood1935} The technique tends to greatly reduce the dimensionality of important integrals, though at a cost of an additional integration over $\lambda$. 

When this field $\phi$ is turned on, the system molecules will redistribute themselves, causing the average local density, $\langle{\hat\rho}_\alpha({\bfr})\rangle$, to deviate from its spatial average $\rho_\alpha=\rho$. What then is  $\langle{\hat\rho}_\alpha({\bfr})\rangle$ for some $\phi_\alpha({\bfr},\lambda)$?

It is helpful to consider this field as changing the constant chemical potential to a spatially varying one, which acts on a site $\alpha$ as 
\begin{equation}
\mu_\alpha(\bfr,\lambda) = \frac{\mu}{N} - \phi_\alpha(\bfr,\lambda).
\label{eq:mu}
\end{equation}
The energy of the coupling of the liquid molecules to this spatially dependent field is then
\begin{equation}
-\sum_\alpha \int d\bfr\ \mu_\alpha(\bfr,\lambda){\hat\rho}_\alpha(\bfr).
\label{eq:muenergy}
\end{equation}

The relevant grand potential is ${\cal G}([\mu],V,T)$ with $[\mu]$ denoting that it is a functional of the field $\mu_\alpha(\bfr,\lambda)$. This potential is related to the grand partition function, $\Xi$, as
\begin{equation}
{\cal G}([\mu],V,T) = - \ln \left [\Xi ([\mu],V,T) \right ].
\label{eq:Omega}
\end{equation}
Within this formalism, it is straightforward to show that the average local density for site $\alpha$ is
\begin{equation}
\rho_\alpha(\bfr,\lambda)\equiv \langle{\hat\rho}_\alpha(\bfr)\rangle_\lambda = - \frac{\delta{\cal G}([\mu],V,T)}{\delta\mu_\alpha(\bfr,\lambda)},
\label{eq:rhoavg}
\end{equation}
where $\delta/\delta \mu_{\alpha}(\bfr,\lambda)$ is a functional derivative with respect to the chemical potential field acting on a site $\alpha$ at position $\bfr$, the field strength held fixed at $\lambda$. Eq.\eqref{eq:rhoavg} is merely a statement that $\rho_\alpha(\bfr,\lambda)$ and $\mu_\alpha(\bfr,\lambda)$ are conjugate to each other.

Now, let $\lambda$ be increased by a small amount $\Delta\lambda$. How would this average local density change? One can perform similar manipulations used to obtain Eq.\eqref{eq:rhoavg} and find
\begin{eqnarray}
&&\frac{d\rho_\alpha(\bfr,\lambda)}{d\lambda} \equiv  {\frac{ \langle {\hat\rho}_\alpha({\bf r})\rangle_{\lambda+\Delta\lambda} -  \langle{\hat\rho}_\alpha({\bf r})\rangle_\lambda}{\Delta\lambda}}\Bigr\vert_{\Delta\lambda\rightarrow 0} \nonumber \\
&&=  - \sum_\beta \int d\bfr' S_{\alpha\beta}(\bfr,\bfr',\lambda) \frac{d\phi_\beta(\bfr',\lambda)}{d\lambda},
\label{drhodlam}
\end{eqnarray}
where $S_{\alpha\beta}(\bfr,\bfr',\lambda)$ is the total density correlation function of the liquid in the external field with strength $\lambda$, and is the analog of Eq.\eqref{eq:Sofr}. Eq.\eqref{drhodlam} is one form of the exact expression for the nonlinear response of the average local density to an external field.

\subsection{Equation for the radial distribution function}
Eq.\eqref{drhodlam} will now be used to obtain an expression for the pair intermolecular correlation function of the bulk liquid. To make further progress, an approximation for $S_{\alpha\beta}(\bfr,\bfr',\lambda)$  is needed. 

Consider first the intramolecular contribution to this function, which is $\rho\omega_{\alpha\beta}(\bfr,\bfr',\lambda)$, where $\omega_{\alpha\beta}(\bfr,\bfr',\lambda)$ is an inhomogeneous generalization of Eq.\eqref{omegaalfbet}. As mentioned above, the field $\phi$ will make the liquid inhomogeneous, attracting a site $\alpha$ to regions where $\phi_\alpha$ is negative and repelling it from regions where this potential is positive. The local density $\rho_\alpha(\bfr,\lambda)$ will then change correspondingly. The change in $\rho\omega_{\alpha\beta}(\bfr,\bfr',\lambda)$ due to $\phi$ will then come from two effects: this change in the local density from its average value $\rho$, and the change in the intramolecular structure of the molecule due to this change in the local density.  Since the interest in this work is liquids at high density, ignore the second effect, which should be less pronounced there due to screening from the other molecules. So,
\begin{equation}
\rho \omega_{\alpha\beta}(\bfr,\bfr',\lambda) \approx \left [\rho_{\alpha}(\bfr,\lambda)\rho_{\beta}(\bfr',\lambda)\right ]^{1/2} \omega_{\alpha\beta}(\bfr-\bfr'),
\label{omegarapprox}
\end{equation}
where $\omega_{\alpha\beta}(\bfr -\bfr')$ is the intramolecular correlation function of a uniform liquid at density $\rho$ and is given by Eq.\eqref{omegaalfbet}. Here, the square root averages the effects of the two sites being at different positions in the field $\phi$. Note that Eq.\eqref{omegarapprox} is exact for atomic liquids.

Next consider the intermolecular contribution to $S_{\alpha\beta}(\bfr,\bfr',\lambda)$, which is $\rho^2 h_{\alpha\beta}(\bfr,\bfr',\lambda)$, where $h_{\alpha\beta}(\bfr,\bfr',\lambda)$ is the inhomogeneous analog of that in Eq. \eqref{Hofrdef}. As for the intramolecular contribution above, the change in it due to the inhomogeneous field $\phi$ will also come from two effects: the change in the local density from its average value $\rho$, and changes in the intermolecular correlations due to this change in local density. Ignore the second effect though, so
\begin{equation}
\rho^2 h_{\alpha\beta}(\bfr,\bfr',\lambda) \approx \rho_{\alpha}(\bfr,\lambda)\rho_{\beta}(\bfr',\lambda) h_{\alpha\beta}(\bfr-\bfr'),
\label{hrapprox}
\end{equation}
where $h_{\alpha\beta}(\bfr-\bfr')$ is the intermolecular correlation function of a uniform liquid at density $\rho$, given by that in Eq.\eqref{Hofrdef}.  
Eq.\eqref{hrapprox} is a variation on the Kirkwood superposition approximation, which is most accurate at low density.\cite{Henderson1988} 

To create a useful form for $S_{\alpha\beta}(\bfr,\bfr',\lambda)$, it is helpful to reconcile the approximations of Eqs.\eqref{omegarapprox}  and \eqref{hrapprox} by letting the latter become
\begin{equation}
\rho^2 h_{\alpha\beta}(\bfr,\bfr',\lambda) \approx \left [\rho_{\alpha}(\bfr,\lambda)\rho_{\beta}(\bfr',\lambda)\right ]^{1/2}\rho h_{\alpha\beta}(\bfr-\bfr').
\label{hrapprox2}
\end{equation}
This form sacrifices the accuracy of Eq.\eqref{hrapprox} at low density, but should be more accurate at high density anyways, which is the interest of this work.
So as in (I): 
\begin{equation}
S_{\alpha\beta}(\bfr,\bfr',\lambda) \approx \frac{1}{\rho}[\rho_\alpha(\bfr,\lambda)]^{1-\eta} [\rho_\beta(\bfr',\lambda)]^{\eta} S_{\alpha\beta}(\vert\bfr-\bfr'\vert),
\label{Sofrapprox}
\end{equation}
where $S_{\alpha\beta}(r)$ is the total density correlation function of a homogeneous liquid, a case of Eq.\eqref{eq:Sofr} above. Given the above arguments and by symmetry, the exponent $\eta = 1/2$, but to derive a closure to PRISM another value will be used.

With Eqs. (\ref{drhodlam}) and (\ref{Sofrapprox}) one finds
\begin{equation}
\frac{dx_\alpha(\bfr,\lambda)}{d\lambda}
=  - \frac{\eta}{\rho}\sum_\beta \int d\bfr' S_{\alpha\beta}(\bfr-\bfr') \frac{d\phi_\beta(\bfr',\lambda)}{d\lambda}\ x_\beta(\bfr',\lambda),
\label{dxdlam}
\end{equation}
where
\begin{equation}
x_\alpha(\bfr,\lambda) = \left[\frac{\rho_\alpha(\bfr,\lambda)}{\rho}\right ]^\eta.
\label{eq:x}
\end{equation}

Now, let the field $\phi$ be due to a molecule of the same type as in the bulk liquid. This molecule is inserted in a fixed configuration $\Re = \{\bfR_1,\bfR_2,\cdots\}$, where $\bfR_1$ is the position of site $1$ on the inserted molecule, etc. So,
\begin{equation}
\phi_\beta(\bfr,\lambda) = \sum_\gamma v_{\beta\gamma}(\bfr - \bfR_\gamma,\lambda),
\label{eq:phi}
\end{equation}
and $v_{\beta\gamma}(r,\lambda)$ is the pair potential between sites $\beta$ and $\gamma$ on different molecules a distance $r$ apart with strength $\lambda$. When $\lambda = 1$, this potential becomes the true potential between these two sites in the bulk liquid. So, $x_\alpha(\bfr,\lambda)\rightarrow x_\alpha(\bfr,\Re,\lambda)$.

Next, with Eq.\eqref{eq:phi}, average Eq.\eqref{dxdlam} over the configurations $\Re$ of the inserted molecule, but hold its site $\zeta$ fixed at $\bfr'$. One obtains a two-point intermolecular correlation function from Eq.\eqref{eq:x}:
\begin{equation}
x_{\alpha\zeta}(\bfr,\bfr',\lambda) = \left\langle x_\alpha(\bfr,\Re, \lambda) V\delta(\bfr' - \bfR_\zeta)\right\rangle_{\Re},
\label{eq:xab}
\end{equation}
with the brackets denoting an average with respect to the intramolecular probability density, ${\cal P}(\Re,\lambda)$, of the inserted molecule, which has normalization
$\int d\Re\ {\cal P}(\Re,\lambda) = 1$. It will be helpful here that no matter the charging strength, $\lambda$, ${\cal P}(\Re,\lambda)$ will be such that the statistical distribution of configurations of the inserted molecule are the same as a molecule in the fully interacting liquid.

Then, upon integrating with respect to $\lambda$ from $\lambda_1$ to $\lambda_2$, Eq.\eqref{dxdlam} becomes
\begin{eqnarray}
x_{\alpha\zeta}(\bfr -\bfr', && \lambda_2) =  x_{\alpha\zeta}(\bfr -\bfr',\lambda_1) -  \frac{\eta}{\rho}\sum_{\beta} \int d\bfr_1 \nonumber \\
\times && S_{\alpha\beta}(\bfr-\bfr_1) \Psi_{\beta\zeta}(\bfr_1-\bfr',\lambda_1,\lambda_2),
\label{dxdlam2}
\end{eqnarray}
where $x_{\alpha\zeta}(r,0) = 1$, and
\begin{eqnarray}
\Psi_{\beta\zeta}(\bfr_1-\bfr^{\prime},\lambda_1,\lambda_2) = \sum_{\gamma}\int_{\lambda_1}^{\lambda_2} && d\lambda' \int d\bfr_2\ \frac{dv_{\beta\gamma}(\bfr_1-\bfr_2,\lambda')}{d\lambda'} \nonumber \\
&& \times x^{(3)}_{\beta\gamma\zeta}(\bfr_1, \bfr_2,\bfr^{\prime},\lambda'),
\label{eq:Psi}
\end{eqnarray}
with the three-point function
\begin{eqnarray}
x^{(3)}_{\beta\gamma\zeta}&&(\bfr_1,\bfr_2,\bfr^{\prime},\lambda) = \nonumber \\
&& \left \langle x_\beta(\bfr_1,\Re,\lambda) V\delta(\bfr_2 - \bfR_\gamma)\delta(\bfr^{\prime} - \bfR_\zeta) \right \rangle_\Re.
\label{eq:x3}
\end{eqnarray}

In the limit $\vert \bfr_1 - \bfr_2\vert \rightarrow\infty$ (or $\vert \bfr_1 - \bfr^\prime\vert \rightarrow\infty$), Eq.\eqref{eq:x3} reduces to $\omega_{\gamma\zeta}(\bfr_2-\bfr^{\prime},\lambda)$, the intramolecular probability density between sites $\gamma$ and $\zeta$ on the inserted molecule at potential strength $\lambda$. With the above stipulation on the form of ${\cal P}(\Re,\lambda)$, this function is the same as that for a molecule in the fully interacting liquid, and so is given by Eq.\eqref{omegaalfbet}. A similar three-point function appears in the expression for the virial pressure for flexible polymers with fixed bond lengths.\cite{Honnell1987}

For the case $\eta = 1$ and $\lambda = 1$, Eq.\eqref{eq:xab} is an alternative but equivalent definition for the radial distribution function of a homogeneous liquid as noted by Percus.\cite{Hansen1986} Given that, for any $\eta$ let
\begin{equation}
g_{\alpha\beta}(r) \approx x_{\alpha\beta}(r,1)^{1/\eta}
\label{eq:gPercus}
\end{equation}
as the pair radial distribution function of the homogeneous liquid. Eqs.\eqref{dxdlam2}-\eqref{eq:gPercus} were the main result of (I).

Note that the three-point function, Eq.\eqref{eq:x3}, contains physical effects that are beyond those modeled by the standard atomic closures to the PRISM equation, intramolecular effects in particular. The intent of the next section is to draw out the dominant ones in a simple manner.

\begin{figure}
\includegraphics[scale=0.37,trim= 0.9in 0.7in 0.0in 0.7in]{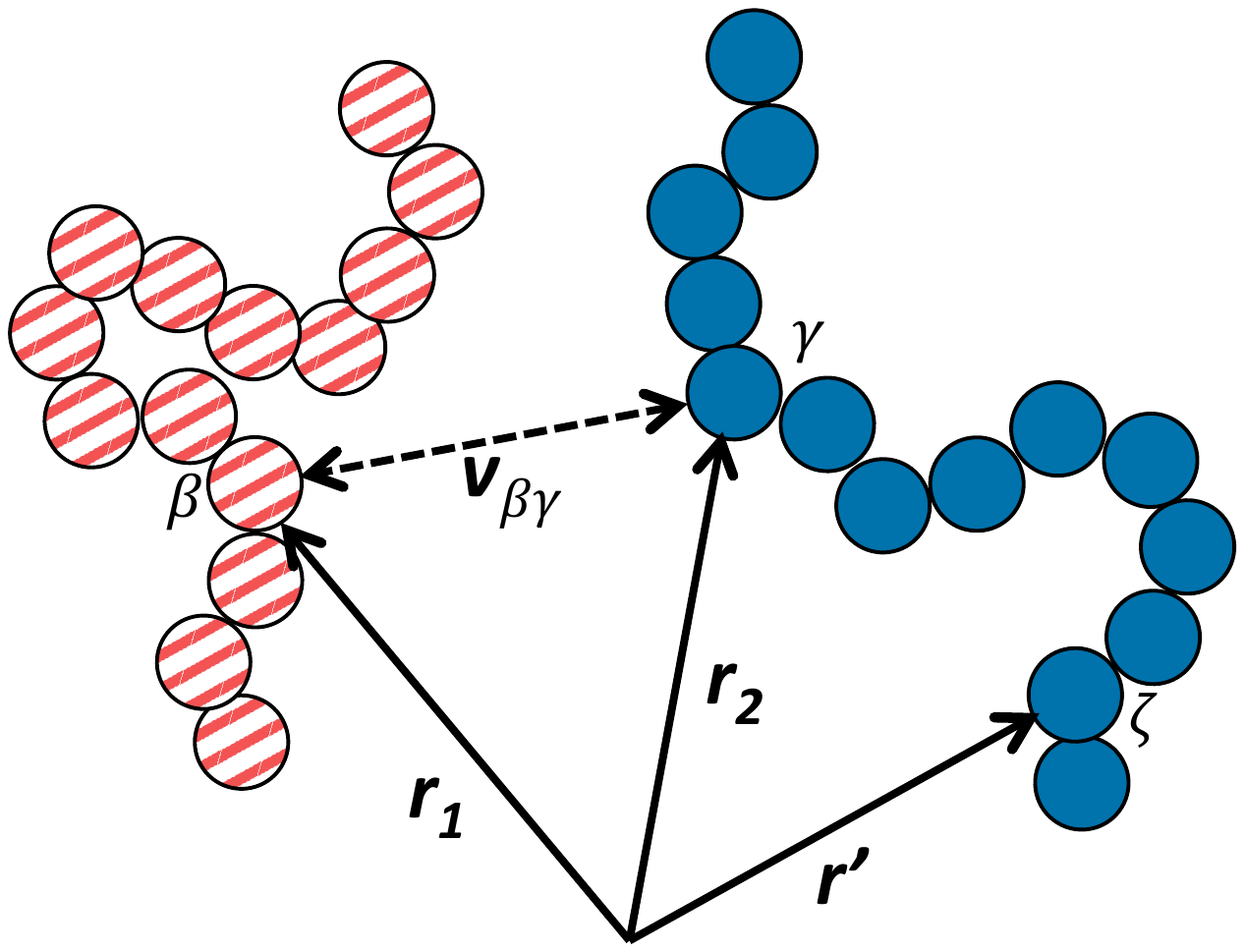}
\caption{\label{figDiagram} Pictorial representation of the sites, their coordinates, and the potential contained in the definitions of $\Psi_{\beta\zeta}$ and $x^{(3)}_{\beta\gamma\zeta}$, Eqs.~\eqref{eq:Psi} and \eqref{eq:x3}, respectively.  For simplicity, the molecules are shown as linear chains. Sites of the inserted molecule are solid blue, and those of a molecule in the liquid are striped red.}
\end{figure}

\subsection{\label{sec:prismClosure}Closure to PRISM}
As stated above, the aim of this work is to derive a closure to the PRISM equation. To that end, set $\eta = 1$ so that Eq.\eqref{eq:gPercus} gives $g_{\alpha\beta}(r) = x_{\alpha\beta}(r,1)$. Let the intermolecular site-site potential, $v_{\alpha\beta}(r)$, be the sum of a short-range, hard core and a long-range, weak tail. This potential depends only on the type of sites, so,
\begin{equation}
v_{\alpha\beta}(r) \equiv v_{kk'}(r)\vert_{\substack{ \alpha \in k \\ \beta \in k'} }= v_{kk'}^{hc}(r) + v_{kk'}^{tail}(r),
\label{vofr}
\end{equation}
where $v^{hc}_{kk'}(r) =\infty$, for  $r < d_{kk'}$ and zero otherwise, and $v^{tail}_{\alpha\beta}(r)=0$ for $r < d_{kk'}$. Restrict the liquid to high, melt densities, so that the radial distribution functions are close to unity outside the core, $r > d_{kk'}$. 

In the following, it will be helpful to imagine that a Kirkwood superposition approximation\cite{Hansen1986,SC88} to the three-point function, Eq.\eqref{eq:x3}, will be sufficient:
\begin{eqnarray}
x^{(3)}_{\beta\gamma\zeta}(\bfr_1,\bfr_2,\bfr^{\prime},\lambda) \approx~ && g_{\beta\zeta}(\bfr_1 - \bfr',\lambda) g_{\beta\gamma}(\bfr_1 - \bfr_2,\lambda) \nonumber \\
\times && \omega_{\gamma\zeta}(\bfr_2-\bfr^{\prime}),
\label{eq:x3approx1}
\end{eqnarray}
where $g_{\beta\zeta}(\bfr_1 - \bfr',\lambda)$ denotes the radial distribution function between a site $\beta$ in the bulk liquid at position $\bfr_1$ and site $\zeta$ on the inserted molecule at position $\bfr'$ with charging strength $\lambda$, etc. This approximation is not universally accurate as it neglects explicit three-body orientational correlations,\cite{SC88} but the charging path taken below for Eq.\eqref{eq:Psi} will allow us to skirt conditions for which they are important.

For the full charging path, let the intermolecular potentials between the sites of the inserted molecule and those of the surrounding liquid molecules be turned on in two steps: first the hard-core with the tail off, then second the tail with the hard-core already on, with $\lambda_x$ denoting the charging value that separates the two steps. Then with Eq.\eqref{eq:Psi} one can define 
\begin{eqnarray}
&&\Psi^{hc}_{\beta\zeta}(r) \equiv  \Psi_{\beta\zeta}(r,0,\lambda_x), \nonumber \\
&&\Psi^{tail}_{\beta\zeta}(r) \equiv  \Psi_{\beta\zeta}(r,\lambda_x,1), 
\label{eq:PsiCases}
\end{eqnarray}
as the contributions from the first and second steps, respectively. The aim here then is first to compute $x_{\alpha\zeta}(r,\lambda_x)$ from Eq.\eqref{dxdlam2} using $\Psi^{hc}_{\beta\zeta}(r)$, and then compute the fully interacting $x_{\alpha\zeta}(r,1)$ using that and  $\Psi^{tail}_{\beta\zeta}(r)$.

First, turn on the hard-core potentials and consider $\Psi^{hc}_{\beta\zeta}$. 
In this case one can expect that the most important intermolecular correlation in $x^{(3)}_{\beta\gamma\zeta}(\bfr_1,\bfr_2,\bfr^{\prime},\lambda)$ in Eq.\eqref{eq:Psi} will be between sites $\beta$ and $\gamma$ as it will strongly moderate the hard-core potential. See Figure~\ref{figDiagram}. The intermolecular dependence of this three-point function on $\bfr^\prime$ can then be ignored and so $g_{\beta\zeta}(\bfr_1 - \bfr',\lambda)$ in Eq.\eqref{eq:x3approx1} can be set to 1. As such,
\begin{equation}
x^{(3)}_{\beta\gamma\zeta}(\bfr_1,\bfr_2,\bfr^{\prime},\lambda) \approx 
g^{hc}_{\beta\gamma}(\bfr_1 - \bfr_2,\lambda) \omega_{\gamma\zeta}(\bfr_2-\bfr^{\prime}),
\label{eq:x3approx}
\end{equation}
where $g^{hc}_{\beta\gamma}(r,\lambda)$ denotes the radial distribution function at the hard-core charging strength $\lambda$.
One then finds,
\begin{equation}
\Psi^{hc}_{\beta\zeta}(r) \approx \sum_\gamma\left [\int_0^{\lambda_x} d\lambda \frac{dv^{hc}_{\beta\gamma}}{d\lambda}g^{hc}_{\beta\gamma}\right ]\ast\omega_{\gamma\zeta}(r),
\label{eq:Psihc}
\end{equation}
where the asterisk ($\ast$) denotes a convolution. The bracket term in Eq.\eqref{eq:Psihc} will be zero beyond the range of the hard-core, and if it were very accurate, $\Psi^{hc}_{\beta\zeta}$ would enable the core condition, $g_{\alpha\beta}(r) = 0$ for $r<d_{kk'}\vert_{\alpha\in k,\beta\in k'}$, to be satisfied in Eq.\eqref{dxdlam2}. Since one cannot necessarily expect the latter condition to hold exactly, replace this bracket term by an effective potential, $-c^{hc}_{\beta\gamma}(r)$, that has the same range as $v^{hc}_{\beta\gamma}(r)$, but is optimized so that the core condition is enforced\cite{Chandler1972}. 

Substituting Eq.\eqref{eq:Psihc} into Eq.\eqref{dxdlam2} and averaging over sites of the same type, one obtains a PRISM equation:
\begin{equation}
H^{hc}_{kk'}(r) = \sum_{l,l'} \Omega_{kl}\ast c^{hc}_{ll'}*S_{l'k'}(r),
\label{eq:H0}
\end{equation}
where $H^{hc}_{kk'}(r) \equiv \rho_k\rho_{k'} [g^{hc}_{kk'}(r) -1]$, with $g^{hc}_{kk'}(r)$ being the radial distribution function between the inserted hard-core molecule and the molecules in the fully interacting liquid. Also, the intramolecular correlation function $\Omega_{kk'}(r)$ is given by Eq.\eqref{omegadef}. The closure has the familiar hard-core Percus-Yevick (PY) form:\cite{Hansen1986,Chandler1972}
\begin{eqnarray}
 &&c^{hc}_{kk'}(r) = 0,\  r > d_{kk'},  \nonumber \\
&&g^{hc}_{kk'}(r) = 0,\  r < d_{kk'}.
\label{eq:H0Closure}
\end{eqnarray}
Note though that Eqs.\eqref{eq:H0} and \eqref{eq:H0Closure} are solved with the fully interacting $S_{l'k'}(r)$, which is held constant.

Next, turn on the tail potentials and consider $\Psi^{tail}_{\beta\zeta}$. Since the tail potential is zero inside the core, and weak and long-ranged, one can expect that the important intermolecular correlation in $x^{(3)}_{\beta\gamma\zeta}(\bfr_1, \bfr_2,\bfr^\prime,\lambda)$ will be between the sites $\beta$ and $\zeta$, the function dropping to zero when $\vert \bfr_1 - \bfr' \vert < d_{kk'}\vert_{\beta \in k,\zeta\in k'}$. See Figure~\ref{figDiagram}. The intermolecular dependence of this function on $ \bfr_2$ can then be ignored, and so $g_{\beta\gamma}(\bfr_1 - \bfr_2,\lambda)$ in Eq.\eqref{eq:x3approx1} can be set to 1. As such, 
\begin{equation}
x^{(3)}_{\beta\gamma\zeta}(\bfr_1,\bfr_2,\bfr^{\prime},\lambda) \approx 
g^{tail}_{\beta\zeta}(\bfr_1 - \bfr^{\prime},\lambda) \omega_{\gamma\zeta}(\bfr_2-\bfr^{\prime}),
\label{eq:x3approxtail}
\end{equation}
where $g^{tail}_{\beta\zeta}(r,\lambda)$ denotes the radial distribution function at the tail charging strength $\lambda$. And so,
\begin{equation}
\Psi^{tail}_{\beta\zeta}(r) \approx \int_{\lambda_x}^1d\lambda \ g^{tail}_{\beta\zeta}(r,\lambda)\sum_\gamma\left [ \frac{dv^{tail}_{\beta\gamma}}{d\lambda}\ast\omega_{\gamma\zeta}(r)\right ].
\label{eq:Psitail}
\end{equation}

In this expression, $g^{tail}_{\beta\zeta}(r,\lambda)$ can be set to 1 outside the core at melt densities, and its dependence on $\lambda$ inside the core can be ignored. Then the charging integration over the tail potential yields just itself at $\lambda = 1$. While $\Psi^{tail}_{\beta\zeta}$ then is formally zero inside the core, that behavior is not expected to be enough to enforce the core condition exactly on the radial distribution function in Eq.\eqref{dxdlam2}. So, replace all of Eq.\eqref{eq:Psitail} with an effective potential $-{\tilde C}^{tail}_{\beta\zeta}(r)$ that equals Eq.\eqref{eq:Psitail} outside the core, but is optimized inside.

Combining Eqs.\eqref{dxdlam2}-\eqref{eq:Psitail}, making the above replacements, and averaging over sites of the same type, yields another PRISM equation:
\begin{eqnarray}
H_{kk'}(r) =~ && H^{hc}_{kk'}(r) + \sum_{l} {\tilde C}^{tail}_{kl}*S_{lk'}(r), \nonumber \\
=~ &&  \sum_{l} {\tilde C}_{kl}*S_{lk'}(r),
\label{eq:Hfull}
\end{eqnarray}
where ${\tilde C}_{kk'} (r)\equiv  \sum_l\Omega_{kl}\ast c^{hc}_{lk'}(r) +  {\tilde C}^{tail}_{kk'}(r)$ using Eq.\eqref{eq:H0}.  Also, $H_{kk'}(r) \equiv \rho_k\rho_{k'} [g_{kk'}(r) -1]$, where $g_{kk'}(r)$ is the radial distribution function between the inserted molecule and the other molecules in the liquid at $\lambda=1$, which is the same as that between the molecules in the fully interacting liquid. 

Optimizing ${\tilde C}^{tail}_{kk'}(r)$ inside the core is the same as optimizing ${\tilde C}_{kk'} (r)$ there, so the closure to Eq.\eqref{eq:Hfull} is then
\begin{eqnarray}
 &&{\tilde C}_{kk'}(r) = \sum_l \Omega_{kl}\ast \left [c^{hc}_{lk'}-v^{tail}_{lk'}\right ](r),\  r > d_{kk'},  \nonumber \\
&&g_{kk'}(r) = 0,\  r < d_{kk'}. 
\label{eq:HfullClosure}
\end{eqnarray}
Since Eqs.\eqref{eq:Hfull} and \eqref{eq:HfullClosure} are for the fully interacting liquid, $S_{kk'}(r)$ can then be obtained self-consistently. 

Eqs.\eqref{eq:H0}, \eqref{eq:H0Closure}, \eqref{eq:Hfull} and \eqref{eq:HfullClosure} form a closed set for the radial distribution functions of a homogeneous liquid. While derived for a single component liquid, they  apply also to a multi-component one.
Given the similarity of the closure Eq.\eqref{eq:HfullClosure} to the MSA, it will be convenient for the rest of this work to refer to the theory above as PRISM with an ``OCMSA" closure, with ``OC" standing for $\Omega*c$. It can be shown that this theory behaves properly in the united atom limit,\cite{CMS86b} the violation of which tends to doom closures at low density. 

This closure will not always have a solution though. The cause lies in Eqs.\eqref{eq:H0} and \eqref{eq:H0Closure}, which describe a hard-core molecule immersed in a fully interacting liquid. If the liquid is cohesive and the temperature is low enough, the surface tension between the inserted hard-core molecule and the liquid may act to expel the inserted molecule. Since the inserted molecule is really the same as any other molecule of that type in the liquid, this problem is an artifact of the two-step charging process above.

Only Lennard-Jones liquids will be examined in this work, and it is well known that the local structure of such liquids at melt densities is dominated by the short-range, essentially hard-core repulsions. The simplest remedy then is to make a reference approximation to the OCMSA closure, similar to that done by Schweizer and Yethiraj (SY)\cite{Schweizer1993,Yethiraj1993} and other researchers.\cite{Hansen1986} That is, in Eq.\eqref{eq:H0}, one lets $S_{l'k'}(r)\rightarrow S^{hc}_{l'k'}(r)$, so that Eqs.\eqref{eq:H0} and \eqref{eq:H0Closure} now describe a liquid of hard-core molecules. Denote this closure  as ``R-OCMSA".

These closures are similar to the reference ``molecular" MSA  (R-MMSA) one suggested by SY:\cite{Schweizer1993}
\begin{eqnarray}
 &&C^{sy}_{kk'}(r) = \sum_{l l'}\Omega_{kl}\ast \left [c^{(0)}_{ll'}-v^{tail}_{ll'}\right ]*\Omega_{l'k'}(r),\  r > d_{kk'},  \nonumber \\
&&g_{kk'}(r) = 0,\  r < d_{kk'}, 
\label{eq:HmmsaClosure}
\end{eqnarray}
which closes a PRISM equation
\begin{equation}
H_{kk'}(r) = \sum_{ll'} C^{sy}_{kl}*\Omega_{ll'}^{-1}*S_{l'k'}(r).
\label{eq:HmmsaEq}
\end{equation}
Here, $c^{(0)}_{kk'}(r)$ is the direct correlation function of a reference liquid of hard-core molecules, and $\Omega_{kk'}^{-1}(r)$ is the functional and matrix inverse of $\Omega_{kk'}(r)$, i.e., 
\begin{equation}
\sum_{l}\int d\bfr^{\prime\prime} \Omega_{kl}(\bfr-\bfr^{\prime\prime})\Omega_{lk'}^{-1}(\bfr^{\prime\prime}-\bfr^{\prime}) = \delta(\bfr-\bfr^\prime)\delta_{kk'},
\end{equation} 
where $\delta_{kk'}$ is the Kronecker delta.

There are one or two differences though. The OCMSA and R-OCMSA closures for the tail effective potential, Eq.\eqref{eq:HfullClosure}, involve an average over the conformations of only the inserted molecule, while as can seen from Eq.\eqref{eq:HmmsaClosure}, the SY closures, similar to the original RISM closure derived by Chandler,\cite{Chandler1973,Ohba1986} have an additional conformational average making them symmetric. In addition, the portion of the OCMSA closure that models the insertion of a hard-core molecule, Eq.\eqref{eq:H0}, is not a reference one, i.e., a hard-core molecule in a liquid of hard-core molecules, but a hard-core molecule in a fully interacting liquid.

Note that in the approximation that $x^{(3)}_{\beta\gamma\zeta}$ is either zero or one at melt densities, and given that the tail potential, $v^{tail}_{\beta\zeta}$, is zero inside the core, the only possible intermolecular correlations that matter in $x^{(3)}_{\beta\gamma\zeta}$ during the charging of the tail potential are between sites $\beta$ and $\zeta$. This necessitates that the hard-core condition be enforced in $\Psi^{tail}_{\beta\zeta}$ by optimization after the intramolecular average of the tail potential. The closure then must contain intramolecular correlations, making it a ``molecular'' one a' la SY.\cite{Schweizer1993}

\section{Numerical Solution}
With known site-site intermolecular potentials, $v_{kk'}(r)$, and intramolecular structure factors, ${\hat\Omega}_{kk'}(q)$, the above equations were solved numerically by iteration on a grid of $N_r$ points with spacing $\Delta r$ in real space or $\Delta q=\pi/(N_r\Delta r)$ in reciprocal space. In this work, $\Delta r$ was set to $0.025$ or $0.05~\AA$, and $N_r$ was set to $2^{16}$.

The algorithm to solve the theory with the R-OCMSA closure was as follows. A guess for $c_{kk'}(r)$ began the algorithm. This was used to obtained an initial value for ${\tilde C}_{kk'}(r) = \sum_l{\Omega}_{kl}*{c}_{lk'}(r)$ for $r<d_{kk'}$, while the closure, Eq.\eqref{eq:HfullClosure}, gave the value for $r > d_{kk'}$, with $c_{kk'}^{hc}(r)$ being obtained from solution of the PRISM equation for hard-core molecules. The PRISM equation and Fourier transform techniques were then used to obtain $g_{kk'}(r)$ from ${\tilde C}_{kk'}(r)$. In analogy with what was implemented in Donley, Heine and Wu,\cite{Donley05a} a measure of the error was defined as $\gamma_{kk'}(r) = \sum_l \rho_k\rho_l g^<_{k l}*\Omega^{-1}_{l k'}(r)$, where $g^<_{kk'}(r) = g_{kk'}(r)$ for $r<d_{kk'}$ and zero otherwise, and $\Omega_{kk'}^{-1}(r)$ is the functional inverse of $\Omega_{kk'}(r)$. The change in ${\tilde C}_{kk'}(r)$ from its last iteration value was identified with the quasi-Newton-Raphson expression $\delta{\tilde C}_{kk'}(r) = - \gamma_{kk'}(r)$ for $r < d_{kk'}$ and zero otherwise. Instead of the standard Picard, the modified method of direct inversion in iterative subspace (MDIIS) of Kovalenko, Ten-no and Hirata\cite{Kovalenko99a,Ishizuka2012} was used to mix this change $\delta{\tilde C}_{kk'}(r)$ with the input values of ${\tilde C}_{kk'}(r)$ from this and previous iterations. 

For the OCMSA closure, the procedure was similar to the above, except $c_{kk'}^{hc}(r)$ was determined self-consistently. It was found efficient to iterate Eqs.\eqref{eq:H0} and \eqref{eq:H0Closure} to convergence for each iteration of Eqs.\eqref{eq:Hfull} and \eqref{eq:HfullClosure}. 

PRISM with the atomic MSA closure and the R-MMSA closure were also solved in most cases. These solutions were obtained in a manner similar to that above, except that the definition of the error measure, $\gamma_{kk'}(r)$, differed given that the  MSA and R-MMSA closures involve $c_{kk'}(r)$ and $C_{kk'}^{sy}(r)$,\cite{Yethiraj1993} respectively.

\section{Results}
In this section, predictions of PRISM theory with the above closures are given for various alkanes, and  binary homopolymer blends of freely-jointed chains and isotopic polyethylene (PE). The main quantity of interest here is the critical temperature, $T_c$, be it the liquid-gas one for the small alkanes or the unmixing one for the blends.

\subsection{\label{sec:symblend}Symmetric blend of freely-jointed chains}
Consider first a binary blend of freely-jointed chains. Denote the binary blend components as $A$ and $B$. Let the chains be structurally symmetric, so $N_A = N_B \equiv N$ and the bond lengths, $b_A = b_B\equiv b$. The structure factor of a linear and overlapping freely-jointed chain, which depends only upon $N$ and $b$, is well known and given in, e.g., Schweizer and Curro.\cite{SC88a}  For simplicity, the site-site potentials were chosen to be shifted Lennard-Jones:\cite{David1994}
\begin{equation}
v_{kk'}(r) = 
  \begin{cases} 
  \infty, & r < d_{kk'} \\
  \epsilon_{kk'}\left [ (\frac{\sigma_{kk'}}{r})^{12} - 2 (\frac{\sigma_{kk'}}{r})^6\right ], & r > d_{kk'}.
  \end{cases}
\label{eq:vrLJShifted}
\end{equation} 
The hard-core diameter, $d_{kk'}$, was set equal to the lengthscale parameter, $\sigma_{kk'}$, and all had the same value $\sigma$, which was assumed independent of temperature. The bond length $b$ was set to $\sigma$, with $\sigma = 1~\AA$, so that $\Delta r/\sigma = 0.05$. Only on-critical mixtures were considered, so $\rho_A=\rho_B$. Since the chains overlap, the total density must to be increased with increasing $N$ to keep the packing fraction constant.\cite{SC88a} The packing fraction was set to $0.5$ for all results in this section. The energy parameters were set to $\epsilon_{AA} = \epsilon_{BB} = 0$ and $\epsilon_{AB} = -\epsilon$, making the blend completely symmetric. The liquid then is not cohesive, so the OCMSA closure is expected to give a solution for all cases. With this specification the only varying parameters were $N$ and $\epsilon$.

The critical point for phase separation of the $A$-$B$ blend was determined using the susceptibility, aka compressibility, route. The predictions using this route have been the most sensitive to approximations in liquid-state theory.\cite{Schweizer1997} Determining the critical point then involved lowering the temperature, i.e., increasing $\epsilon$, and monitoring the susceptibilities, ${\hat S}_{kk'}(q=0)$. Let ${\Lambda}(q) \equiv {\rm det} ( {\bf\hat\Omega}(q){\bf\hat S}(q)^{-1} )$, where the boldface indicates an $n_t\times n_t$ matrix, and matrix multiplication and inversion are implied. Then for any $N$, the spinodal, which in this case was also the critical point, was considered to have been reached when $\Lambda (0) \leq \xi$, a small number. (The ceiling $\xi$ varied in this work from 0.01 to 1, depending on the system examined.) The value of $\epsilon$ at this point was identified with its critical value, $\epsilon_c$. 

\begin{figure}
\includegraphics[scale=0.47,trim= 0.05in 0.1in 0.0in 0.05in]{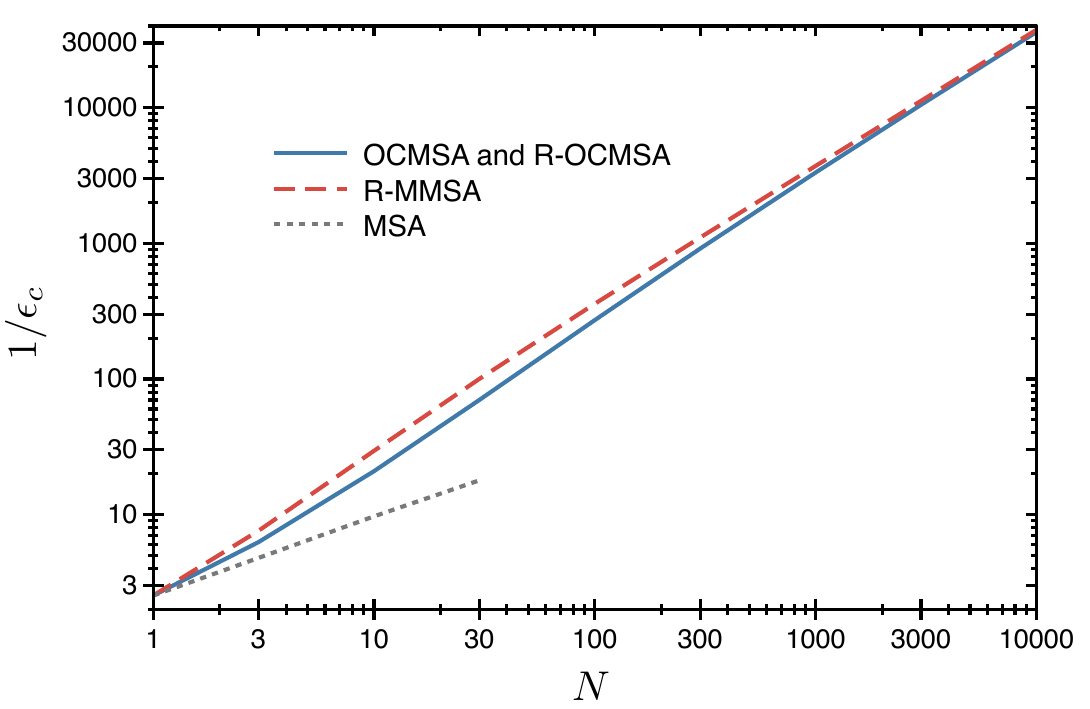}
\caption{\label{figEpscVsN} Inverse of the critical value of the scaled energy parameter, $\epsilon_c$, as a function of chain length, $N$, for a symmetric polymer blend of freely-jointed chains as described in Sec.~\ref{sec:symblend}. The critical temperature, $T_c$, is proportional to $1/\epsilon_c$. PRISM theory predictions using the OCMSA and R-OCMSA (solid blue), R-MMSA (long dashed red) and MSA (short dashed gray) closures are shown.}
\end{figure}

Figure~\ref{figEpscVsN} shows $1/\epsilon_c$ as a function of chain length $N$ for PRISM theory using the OCMSA and R-OCMSA closures. The predictions of these two closures are essentially identical for this system. Since $T_c \sim 1/\epsilon_c$, it can be seen that at large $N$  the critical temperature is proportional to $N^{\nu}$ with $\nu = 1.03 \approx 1$, in agreement with Flory theory. 

As can be also seen, the predictions of the R-MMSA closure for $T_c$ for $N>1$ are higher than that for the OCMSA and R-OCMSA closures, reaching a maximum of about 40\% between $N=10$ and 30. For very large $N$ though, the predictions of the R-MMSA appear to approach asymptotically that of the OCMSA closures, differing by only 4\% for $N=10000$.  As measured by the number of temperature steps needed for $\Lambda (0)$ to stably drop to $\xi$, the difficulty of solving all these closures was about the same for all $N$. The difficulty of determining $T_c$ using the MSA closure though increased rapidly with $N$, with the number of steps increasing almost linearly with $N$. Here, the largest $N$ for which $T_c$ was computed for the MSA closure was 30. At these smaller $N$ for the MSA, $T_c$ was determined to scale with an exponent $\nu \approx 0.57$, in rough agreement with the known asymptotic value of $0.5$.\cite{Schweizer1997}

\subsection{\label{sec:methane}Methane and ethane}
While the focus of this paper is on polymeric liquids, particularly isotopic PE in Sec.~\ref{sec:isoblend} below, it is interesting for comparison to explore the predictions of the theories for two smaller alkanes, methane and ethane.  Examined will be their liquid-gas critical temperature, $T_c$, and critical molecular number density, $\rho_c$, both which have been measured experimentally.\cite{Teja1990}

Methane, $\rm CH_4$, and ethane, $\rm (CH_3)^2$, were modeled in the usual way as a united atom and dimer, respectively. Theory then requires their site-site potentials, each suitably mapped to a short-range hard-core and long-range tail, and the pair intramolecular structure factor of the dimer. The latter is well known,\cite{DCM94} and the bond length was set to $1.54~\AA$ as done for the TraPPE model\cite{MS98} of alkanes. The nonbonded site-site potentials were approximated as Lennard-Jones with the length parameters, $\sigma$, and scaled energy parameters, $\epsilon$, for the $\rm CH_3$ and ${\rm CH_4}$ groups also given by the TraPPE model. Values are shown in Table~\ref{tabPOT}.

\begin{table}
\caption{Lennard-Jones parameters for the alkane united atom sites used in this work. These are from the TraPPE model of Martin and Siepmann.\cite{MS98} For the scaled energy parameter, $\epsilon$, the temperature is in units of Kelvin (recall that all energies in this work are scaled in units of $k_BT$).}
\setlength{\tabcolsep}{20pt}
\begin{tabular}{ccc}
\hline \hline
Site & $\sigma~(\AA)$ & $\epsilon$ \\
\hline
${\rm CH_2}$ & 3.95 & 46$/T$  \\
${\rm CH_3}$ & 3.75 & 98$/T$   \\
${\rm CH_4}$ & 3.73 & 148$/T$  \\
\hline\hline 
\end{tabular}
\label{tabPOT} 
\end{table}

Estimates of the hard-core diameters, $d_{kk'}$, are needed to map each of the above site-site Lennard-Jones potentials onto a hard-core plus tail. The Barker-Henderson and Weeks-Chandler-Andersen mappings have been the most common.\cite{Hansen1986} Here however, the perhaps simpler and more accurate method of range optimization was used.\cite{Donley2004,Donley2015,Hoye2016,Kobryn2016} Applied to Lennard-Jones potentials, the range optimized value for $d_{kk'}$ is the smallest that allows $g_{kk'}(r)$ to be non-negative for all $r$.\cite{Donley2004} It can be shown that it also makes the effective potentials, i.e., direct correlation functions, $c_{kk'}(r)$, continuous at $r=d_{kk'}$. Range optimization plays a similar function for handling interactions in ionic liquids as the Kovalenko-Hirata linearization scheme for the atomic hypernetted-chain (HNC) closure,\cite{Kovalenko99b} but is not tied to any particular theory.\cite{Donley2015} 

In past work modeling polyelectrolyte solutions, range optimization was accomplished by solving the theory with guesses for the hard-core diameters and then varying them until the range optimized criterion was satisfied.\cite{Donley05a} For the alkanes and isotopic PE though it was found that the hard-core diameters varied little with temperature. For these liquids, range optimization was achieved by approximating the potential as hard-core if it exceeded a threshold, $v_{thres}$, at any point $r$. A value for $v_{thres}$ of 1.1, 0.95 and 0.6 was found to work for methane, ethane and PE, respectively, in most cases.

As can be seen in Table \ref{tabResults}, the OZ-MSA prediction for the liquid-gas $T_c$ for methane is about 14\% less than the experimental value, while $\rho_c$ is within 1\%. All the molecular closures considered here, OCMSA, R-OCMSA and R-MMSA, rely more or less on the PRISM-PY solution for a hard-core reference state, which is known to be less accurate at low density for polymers.\cite{CWGW99} As such the predictions of these closures for liquid-gas critical properties should be most accurate in the atomic limit where they all reduce to the MSA. Thus, one can consider this accuracy for methane as the best the PRISM closures could achieve for the liquid-gas critical properties of alkanes, including PE. 

For the dimer ethane, data in Table \ref{tabResults} shows that the predictions of the PRISM closures are in better agreement with experiment for $T_c$ than for methane, being 2-5\% less. Given the discussion above, this very good agreement can be considered fortuitous. As expected, the predictions for $\rho_c$ are not as accurate as for methane, being 6-13\%  higher than the experimental value.

\begin{table*}
\caption{Summary of the comparisons in Secs.~\ref{sec:methane}, \ref{sec:pemelt} and \ref{sec:isoblend} of the various theories with experiment and MD simulation. Definitions of the measured quantities are given in the text. All critical properties are obtained from the susceptibility route. Each theory source is denoted by its closure. The system ``PE" denotes a polyethylene melt with properties stated in Sec.~\ref{sec:pemelt}. The system ``IsoPE" denotes a binary isotopic blend of PE with properties stated in Sec.~\ref{sec:isoblend}. If a numerical solution to a theory was not found, it is denoted by ``n/f".}
\setlength{\tabcolsep}{3pt}
\begin{tabular}{ccccccc}
\hline \hline
System & Source & $T_c~ (K)$ & $\rho_c~(\times 10^{3}~\AA^{-3})$ & $T_s~(K)$ & ${\hat S}(0)/\rho_m$ & ${\tilde u}_{inter}~(\AA^{-3})$  \\
\hline
Methane & Expt\cite{Teja1990} & 191 & $6.1$ &  &  &    \\
& MSA & 165 & $6.1$  &   &  &   \\
Ethane & Expt\cite{Teja1990} & 305 & $4.07$ &   &  &  \\
& MSA & 289 & $4.61$ & & & \\
& OCMSA & 300 & $4.58$ & & & \\
& R-OCMSA & 290 & $4.45$ & & & \\
& R-MMSA & 292 & $4.32$ & & & \\
PE & MD & & & & 0.291 & $-0.032\pm 0.001$ \\
& MSA & & & n/f & 1.13 & $-0.035$ \\
& R-OCMSA & & &  278 & 1.03 & $-0.035$ \\
& R-MMSA & & & 278 & 1.03 & $-0.035$ \\
IsoPE & Expt\cite{Londono1994} & 400 & & & & \\
& MSA & n/f & & & & \\
& R-OCMSA & 500 & & & & \\
& R-MMSA & 505 & & & & \\
\hline\hline 
\end{tabular}
\label{tabResults} 
\end{table*}

\subsection{\label{sec:pemelt}Normal polyethylene melt}
The properties of a melt of linear hydrogenated, i.e., normal, PE are examined here for a single temperature, pressure and chain length. The temperature and pressure were chosen to be $430~K$ and $1~atm$, respectively, which is consistent with an experiment on PE of Zoller\cite{Zoller1979} who measured the average monomer number density to be $\rho_m=0.0334~\AA^{-3}$. For a reason stated in Sec.~\ref{sec:isoblend} below, the chain length $N$ was set to 4373. The statistical segment length measured by Londono et al. at a similar temperature and density was $6\AA$,\cite{Londono1994} which for $N = 4373$ gives an average radius of gyration, $R_g=162\ \AA$.

Theory requires that the equilibrium chain structure factor, ${\hat \Omega}(q)$, and intermolecular potentials be specified. Previous theoretical work used the rotational isomeric state (RIS) model to compute ${\hat\Omega}(q)$,\cite{Honnell1991,Melenkevitz1997} but given that only one case was needed here, it was computed instead by molecular dynamics (MD) simulation. The simulation consisted of 50 PE chains, each with the same number of 4373 monomers. The initial configuration was chosen so that the average $R_g$ was $161\AA$, close to the experimental value mentioned above. As is usual for long chains, end effects were ignored, so that all sites were modeled as united atoms of $\rm CH_2$.The nonbonded potentials were modeled as Lennard-Jones, with values for the length and energy parameters,  $\sigma_{HH}$ and $\epsilon_{HH}$ (the subscript $H$ denoting a $\rm CH_2$ group), respectively, specified by the TraPPE model for $\rm CH_2$ monomers,  given in Table~\ref{tabPOT} above.  The bond angular and torsion potentials were set to be the same as those used by Martin and Siepmann in their study\cite{MS98} of alkanes. The bond stretch potential was modeled as a stiff harmonic, with the average value set to $1.54\ \AA$ appropriate to PE. The LAMMPS simulation package\cite{Plimpton95} was used. The constant $NPT$ simulation was run to 50 {\it ns}, which was not long enough to equilibrate fully the chains, they being much larger than the simulation box size, but long enough to equilibrate the density modes with wavelength less than the box size and the chain structure on those length scales. The average chain $R_g$ changed only slightly then. Averages were taken using data from 40 to 50 {\it ns}. The average $\rho_m$ for the above $T$ and $P$ was measured to be $0.0337\ \AA^{-3}$, which is within 1\% of the experimental value.

Theory further requires that the Lennard-Jones potential between $\rm CH_2$ sites be suitably mapped to a short-range hard-core and long-range tail. The mapping was the same as described in Sec.~\ref{sec:methane} above. No solution using the OCMSA closure was found below $T=600\ K$ at this density.

\begin{figure}
\includegraphics[scale=0.47,trim= 0.05in 0.1in 0.0in 0.05in]{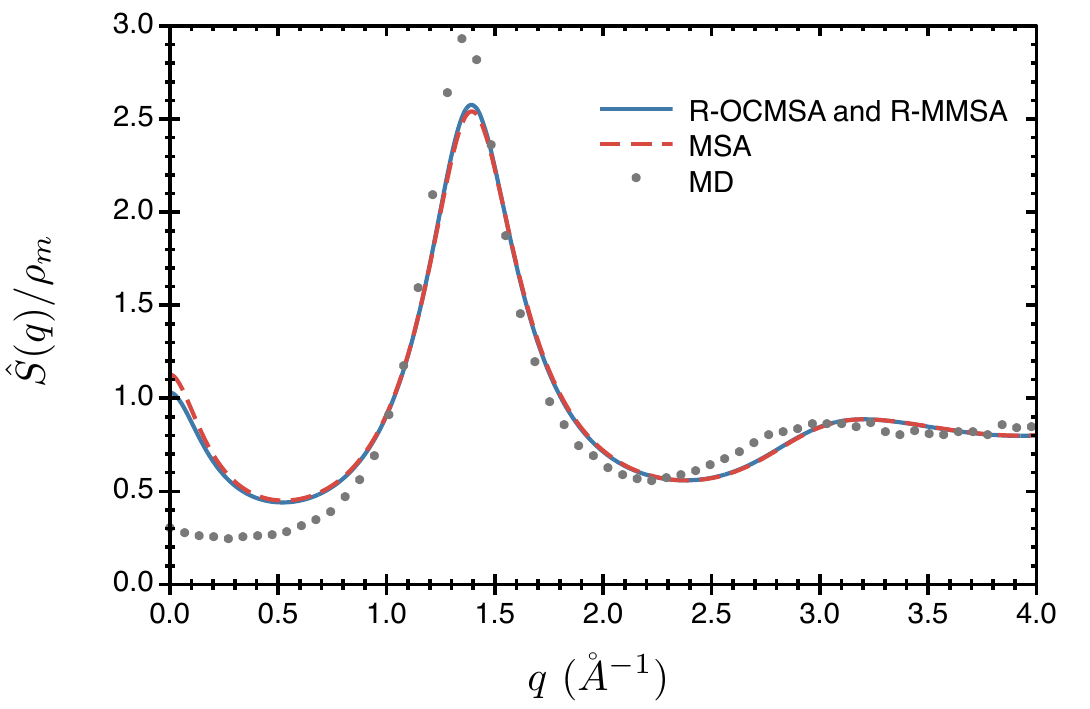}
\caption{\label{figSqPE} Scaled monomer-monomer structure factor, ${\hat S}(q)/\rho_m$, as a function of wavevector $q$ for a melt of hydrogenated, i.e., normal, PE. Conditions are described in Sec.~\ref{sec:pemelt}. The meaning of the symbols is given in the figure legend.}
\end{figure}

Figure~\ref{figSqPE} shows the monomer-monomer structure factor ${\hat S}(q)$ as a function of wavevector $q$ for the PE melt. As can be seen, PRISM with the MSA, R-OCMSA and R-MMSA closures give about the same predictions for ${\hat S}(q)$ for all wavevectors. 
For PE melts of shorter chains, two-molecule theory appears to give better agreement with MD simulation data for local structure, $0.5 < q < 4.0~\AA^{-1}$, but the accuracy of these types of MSA closures in Figure~\ref{figSqPE} for the susceptibility, ${\hat S}(q=0)$, are comparable.\cite{Li2022} See Table~\ref{tabResults}. The overestimation of ${\hat S}(q=0)$ and thus the compressibility is typical for PRISM theories when cohesive liquids are modeled.\cite{CWGW99} A possible explanation for that observed with two-molecule theory is that the HNC-like medium-induced potential in that theory overestimates the compression effect of surrounding chains.\cite{Li2022} A remedy then would be to include ``bridge" corrections in the medium-induced potential.\cite{Hansen1986,Tanaka2014} For the MSA class of PRISM closures here, the cause is not as straightforward, but could be partly due to the product approximation for the density-density correlation function in Eq.~\eqref{Sofrapprox}.

\begin{figure}
\includegraphics[scale=0.47,trim= 0.05in 0.1in 0.0in 0.05in]{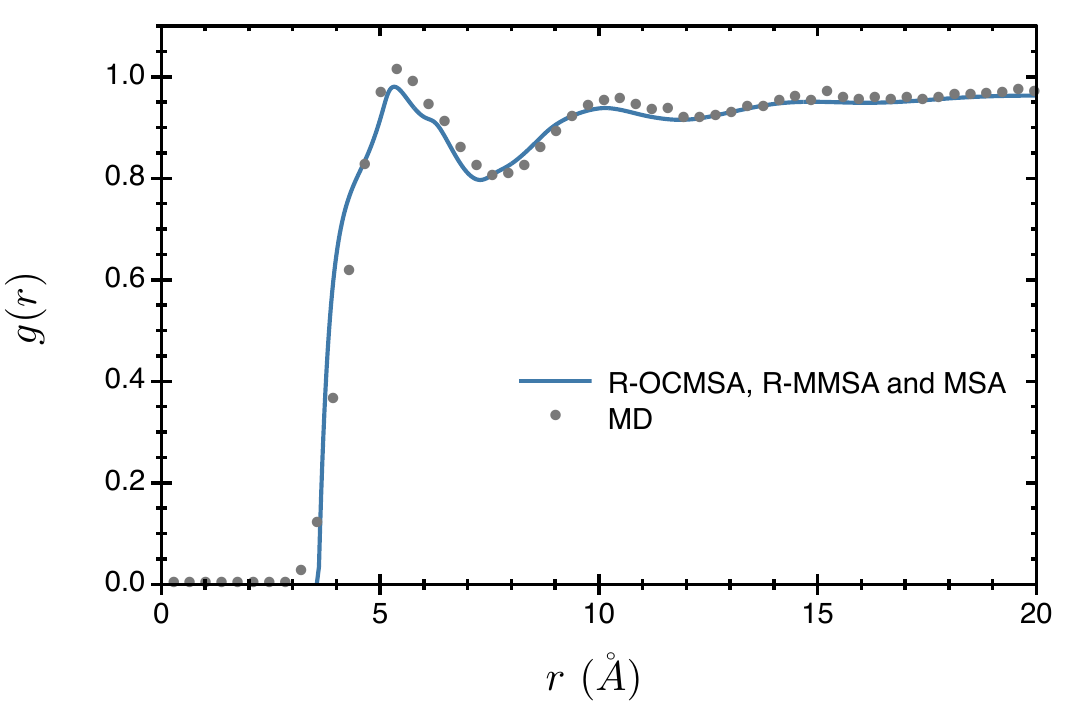}
\caption{\label{figGrPE} Monomer-monomer radial distribution function, $g(r)$, as a function of radial distance $r$ for a melt of normal PE. The conditions are the same as for Figure~\ref{figSqPE}. The meaning of the symbols is given in the figure legend. The predictions of the various closures are essentially the same at these lengthscales and so are shown as a single curve.}
\end{figure}

The radial distribution function, $g(r)$, was also examined for the PE melt. Since the average chain end-to-end distance, $R_{ee}\approx 402~\AA$, was larger than the simulation box width, $L\approx 186~\AA$, extra care was needed to compute this quantity from the MD data.\cite{Pant1993} Figure~\ref{figGrPE} shows $g(r)$. The predictions of all the closures are pretty much identical at these short lengthscales for this system and are thus shown by the same curve. As can be seen, the theoretical predictions are in good agreement with the MD data. The theoretical $g(r)$ has a form midway between the simulation data and a past prediction of hard-core PRISM-PY theory for a similar PE system.\cite{Honnell1991} 

A measure of the cohesiveness of the liquid is the intermolecular contribution to the internal energy per unit volume. For the PE liquid it is defined as\cite{Hansen1986}
\begin{eqnarray}
{\tilde u}_{inter} = && ~\frac{1}{2V}\sum_{i\ne j}\sum_{\alpha\beta} \left \langle u(\vert\bfr_{i\alpha}-\bfr_{j\beta}\vert)\right \rangle \nonumber \\
\simeq&& ~2\pi \rho_m^2 \int_0^\infty dr\ r^2 u(r) g(r),
\label{uinter}
\end{eqnarray}
where $u(r)$ is the intermolecular pair potential between ${\rm CH_2}$ sites. As can be seen from Table~\ref{tabResults}, the theory prediction is in good agreement with the MD value (per particle this energy is about $1~k_BT$). Here, the variance in the MD value is from the two types of smoothing procedures done for the MD data for $g(r)$, the value of ${\tilde u}_{inter}$ being sensitive to the shape of $g(r)$ around the effective hard-core diameter. 

Recall that the OCMSA and R-OCMSA closures were derived with the intermolecular correlations embodied in the three-point function, $x^{(3)}$, defined by  Eq.~\eqref{eq:x3}, approximated as zero if the two chains overlapped and unity if they did not. This translates to $g(r)=\theta (r-d)$, where $\theta(x)$ is the Heaviside step function and $d$ is the hard-core diameter. If that approximation is used in Eq.~\eqref{uinter} with $d=3.575~\AA$ as determined by solution of the PRISM equation with any of the MSA closures, one finds, ${\tilde u}_{inter} = -0.033~\AA^{-3}$. This value agrees better with the MD data than the full theory prediction. While this better agreement can be considered partly fortuitous, it seems to support the validity of the step function approximation to $x^{(3)}$ for melts.

Since $P= 1\ atm$, this thermodynamic state should be close to the liquid-gas coexistence boundary. It is interesting then to determine the spinodal temperature, $T_{s}$, at this density. It was found that PRISM with the R-OCMSA and R-MMSA closures both predict that $T_{s} \simeq 278~K$, which is above the glass transition temperature.\cite{Boyer1973} See Table~\ref{tabResults}. At temperatures below $296 K$, PRISM with the MSA closure developed multiple solutions, so $T_{s}$ was not found. It became progressively more difficult to find a solution using the R-MMSA closure for $T < 450~K$. In particular the sensitivity to the initial guess for $c_{kk'}(r)$ greatly increased. A similar difficulty was encountered in the application of the R-MMSA to diblock copolymers.\cite{David1994} It was unclear though whether these issues at low $T$ with that closure and the MSA were intrinsic to them, or that the algorithm used here could be improved. 

\subsection{\label{sec:isoblend}Isotopic polyethylene blend}
In this section, comparison is made with experimental thermodynamic data from Londono et al.\cite{Londono1994} for a melt binary blend of linear isotopic PE. In particular, the experimental sample of interest had $N_H=4598$, $N_D = 4148$ and $\phi_D=0.457$, where $N_H$ and $N_D$ are the average number of monomers per chain for the hydrogenated (normal) and deuterated component, respectively, and $\phi_D$ is the volume fraction of the deuterated component. For this sample, the critical $\chi$ parameter given their equation 3 was $4.6\times 10^{-4}$, which using their figure 7 the critical temperature was inferred to be approximately $400\ K$.

First, it was found convenient to approximate the above experimental case as one with $N_H=N_D\equiv N=(4598+4148)/2=4373$ and $\phi_D=0.5$. As usual for long chains, end effects were  ignored, so all sites on the hydrogenated and deuterated chains were $\rm CH_2$ and $\rm CD_2$ groups, respectively. As for the alkanes in Secs.~\ref{sec:methane} and \ref{sec:pemelt} above, the nonbonded site-site potentials were approximated as Lennard-Jones. The values for the length and energy parameters between $\rm CH_2$ sites,  $\sigma_{HH}$ and $\epsilon_{HH}$, respectively, are given in Table~\ref{tabPOT}. The volume change upon mixing for the isotopic PE liquid is small\cite{Melenkevitz1997,Melenkevitz2000} and so was ignored. Consequently, $\sigma_{HD}$ and $\sigma_{DD}$ were set equal to  $\sigma_{HH}$. The scaled energy parameter between $\rm CD_2$ groups, $\epsilon_{DD}$, was represented as $\epsilon_{HH}/(1+\delta)^2$. Averaging the four experimental estimates quoted in Bates et al.,\cite{Bates1988} along with the original Bell one,\cite{Bell1942} gives $\delta\approx 0.0161\pm 0.0007$, independent of temperature.  Bertholet scaling was assumed, so $\epsilon_{HD} = (\epsilon_{HH}\epsilon_{DD})^{1/2} = \epsilon_{HH}/(1+\delta)$.

Next, any differences between the intramolecular correlations in the hydrogenated and deuterated liquids were ignored, so ${\hat\Omega}_{DD}(q) = {\hat\Omega}_{HH}(q) \equiv {\hat\Omega}(q)$. While ${\hat\Omega}(q)$ in general varies with density and temperature, in the theory for unmixing of blends the overall density was held fixed and the temperature changes were not that large, so this function was needed for only one density and temperature. Thus, ${\hat\Omega}(q)$ was set to be the same as for the PE melt described in Sec.~\ref{sec:pemelt} above.

For the isotopic blend, it was found that PRISM with the R-OCMSA and R-MMSA closures predicts an unmixing critical temperature, $T_c$, about 25\% higher than the above experimental estimate. See Table~\ref{tabResults}. The critical point using PRISM-MSA was not found, though if it exists it is below $320\ K$. As described in Sec.~\ref{sec:methane} and can be also seen inTable~\ref{tabResults}, the prediction of atomic OZ-MSA theory for $T_c$ for methane is 14\% less than the experimental value. So the predictions for $T_c$ for isotopic PE are not terribly worse, though they overestimate rather than underestimate. This overestimation could be expected from the trend for $T_c$ of ethane versus methane as shown in Table~\ref{tabResults}. 

While this comparison involved no adjustable parameters, some sources of uncertainty for $T_c$ are $\delta$ and the potentials used for hydrogenated PE. The uncertainty in $\delta$ translates in this case to an uncertainty in $T_c$ of about $17\ K$. A better estimate of $\delta$ would thus be helpful, there being an enduring interest in deuterated polymers.\cite{Li2021}

\begin{figure}
\includegraphics[scale=0.47,trim= 0.05in 0.1in 0.0in 0.05in]{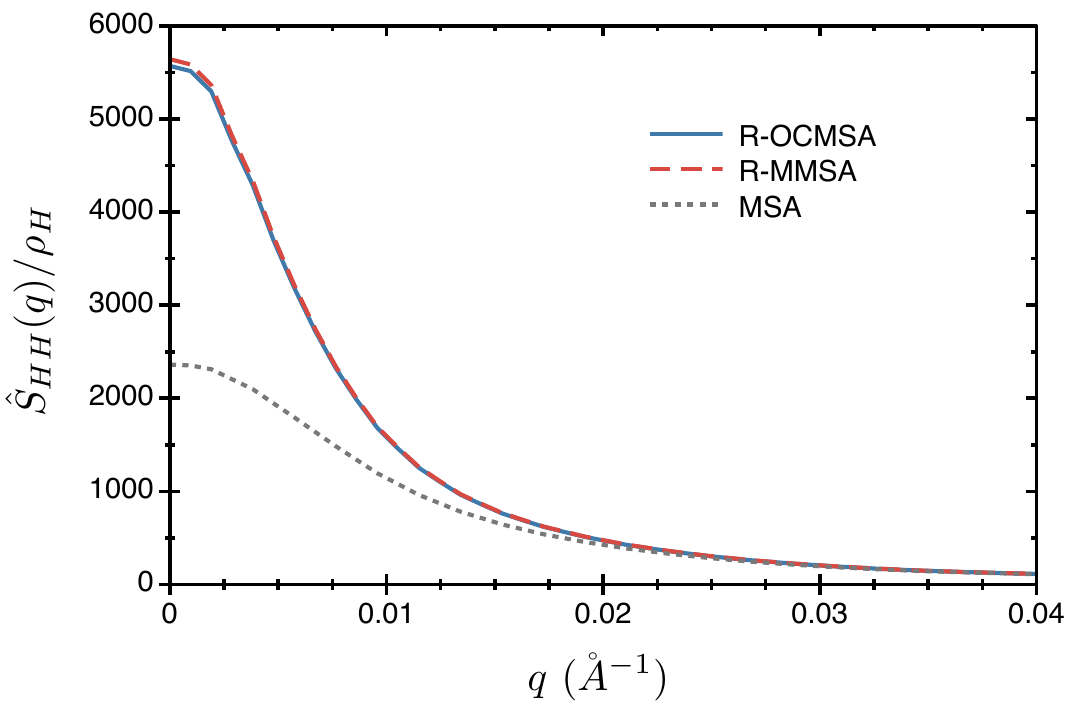}
\caption{\label{figSqHH} Scaled partial structure factor, ${\hat S}_{HH}(q)/\rho_H$, as a function of wavevector $q$ for the isotopic PE blend described in Sec.~\ref{sec:isoblend}. Here, $T=600~K$, so the liquid is about 100 degrees above $T_c$ as predicted by PRISM theory with the R-OCMSA and R-MMSA closures. The meaning of the curves is given in the figure legend.}
\end{figure}

For temperatures at least slightly above these values for $T_c$, the MSA closure gives predictions for the partial structure factors of this binary blend that are about the same as those from the R-OCMSA and R-MMSA closures for wavevectors away from $q = 0$. For example at $T=600~K$, the MSA prediction for ${\hat S}_{HH}(q)/\rho_H$ is essentially identical to that of the R-OMCSA and R-MMSA closures for $q \geq 0.02~\AA^{-1}\approx \pi/R_g$, till diverging at smaller $q$, reaching 2362, 5572 and 5644, respectively, at $q=0$. See Figure~\ref{figSqHH}.

\section{\label{sec:summary} Summary and Discussion}
In summary, building on the previous work in (I),\cite{Donley05b} a simpler derivation of an integral equation for the radial distribution function of molecular liquids in equilibrium was presented.  Relations from nonlinear response theory were used. Further approximations were then made to derive a closure to the PRISM equation appropriate to neutral polymer melts. The resultant closure, ``OCMSA", was found to not always have a solution though. It was argued that the cause was the particular charging path taken in the derivation, it being needed to separate out the effects from the hard-core and tail portions of the site-site potentials. But given the nature of correlations in neutral polymer melts, a reference approximation to this closure could be made. This reference closure, ``R-OCMSA", was found to have solutions for all systems and thermodynamic states studied. The R-OCMSA closure has a form similar to the reference molecular MSA (R-MMSA) closure of Schweizer and Yethiraj (SY), except the latter has an additional intramolecular molecular average on the direct correlation function, making it symmetric.\cite{Schweizer1993} 

For a model homopolymer binary blend of freely-jointed chains, the OCMSA and R-OCMSA closures gave essentially identical predictions for the unmixing critical temperature, $T_c$, from the susceptibility route. This $T_c$ was determined to scale linearly with molecular weight in agreement with Flory theory. The prediction for $T_c$ of these closures differed greatest from that of the R-MMSA for intermediate chain lengths, $10\leq N\leq 30$, with the latter closure about 40\% higher there. At the largest chain lengths though, the closures appeared to approach the same value with the difference for $N=10000$ being only 4\%. 

In comparison with experimental data for isotopic polyethylene blends, the R-OCMSA closure predicted a $T_c$ about 25\% too high. The R-MMSA closure gave only a slightly higher value for $T_c$, which is consistent with that seen for the freely-jointed chain blend in the limit of large molecular weight. Both these values reflect an accuracy only moderately worse than the predictions of Ornstein-Zernike-MSA theory for neutral atomic liquids, at least for the one case examined.

With the exception of the melt blend of freely-jointed chains at intermediate chain lengths, the R-OCMSA and R-MMSA closures gave about the same predictions for all cases of liquids of homogenous molecules examined. The ease of finding a numerical solution sometimes varied between them though.

The derivation given in Sec.~\ref{sec:prismClosure} shows that PRISM-MSA is not the most sound theoretical description of neutral polymer melts. That is, for such a liquid, intramolecular correlations necessarily appear in a closure to the PRISM equation.\cite{Schweizer1993} It can be argued that RISM-MSA, whether viewed as an optimized RPA theory\cite{Chandler1972} or derived from a Gaussian field theory,\cite{Chandler1993} incorporates the effect of the short-range, hard-core repulsion portion of potentials in a first order manner beyond the RPA. If so, then the derivation here (and the earlier work of SY) shows that achieving proper scaling for $T_c$ via the susceptibility route for a polymer blend within PRISM can be attained when the hard-core condition is enforced, but only if the feedback of that onto the original contribution from the long-range tail is treated properly, which is a second order effect. As mentioned in (I), all theories are approximate, and since liquid-state theories by their nature concentrate on local structure, it should not be surprising, at least in hindsight, that extra effort has been needed to get PRISM theory to model properly such a long wavelength property.\cite{Pini98}

In spite of this limitation of the MSA closure, it was shown here that if the thermodynamic state was at least moderately distant from a second order phase transition boundary, the predictions of this closure for liquid structure were very similar to that of the R-OCMSA and R-MMSA, especially for wavevectors $q > 0$. 
There is not a large body of research comparing PRISM-MSA with simulation or experiment.\cite{Schweizer1997} An example though from our past work was explaining the invariance of density correlations seen in experiments of strongly charged polyelectrolyte solutions\cite{Essafi99} as primarily due to liquid-state effects, rather than the conventional explanation of counterion condensation.\cite{Donley2004,Donley06} In this case, this success can viewed as at least partly due to RISM/PRISM originating as an optimized RPA theory.\cite{Chandler1972} To my knowledge no theory of classical ionic liquids to date predicts a second order phase transition from the susceptibility route.\cite{Aqua2005,Shen2018} 
Yet, it still would be interesting to capture this behavior of strongly charged polyelectrolytes in a molecular version of the MSA closure, such as the OCMSA or similar.

\begin{acknowledgments}
I thank Marina Guenza for helpful discussions.
\end{acknowledgments}

\section*{Author Declarations}
The author has no conflicts to disclose.

%


\bibliography{nlr_ocRef}


%
\end{document}